\documentclass[]{iopart} 

\usepackage{color}
\usepackage{graphics}
\usepackage{graphicx}
\usepackage{amssymb}
\usepackage{iopams}
\usepackage{color}
\usepackage{iopams}

\newcommand{\text}[1]{\mathrm{#1}}

\begin{document}
\title{Unveiling a nematic quantum critical point in multi-orbital
  systems}
\author{Christoph M. Puetter$^{1}$, Sylvia D. Swiecicki$^{1}$ and Hae-Young Kee$^{1,2}$}

\address{$^1$ Department of Physics, University of Toronto, Toronto,
  Ontario M5S 1A7, Canada}
\address{$^2$ Canadian Institute for Advanced Research, 
  Quantum Materials Program, Toronto, Ontario M5G 1Z8, Canada}

\ead{cpuetter@comas.frsc.tsukuba.ac.jp, hykee@physics.utoronto.ca}

\begin{abstract}
Electronic nematicity, proposed to exist in a number of transition 
metal materials, can have different microscopic origins.
In particular, the anisotropic resistivity and meta-magnetic jumps 
observed in Sr$_{3}$Ru$_{2}$O$_{7}$
are consistent with an earlier proposal that the
isotropic-nematic transition is generically first order 
and accompanied by meta-magnetism when tuned by a 
magnetic field.
However, additional striking experimental features
such as a non-Fermi liquid resistivity and critical thermodynamic behaviour
imply the presence of an unidentified quantum critical point (QCP).
Here we show that orbital degrees of freedom play an essential role in 
revealing a nematic QCP, even though it is overshadowed by a nearby 
meta-nematic transition at low temperature.
We further present a finite temperature phase diagram including the 
entropy landscape and discuss our findings in light of the phenomena 
observed in Sr$_{3}$Ru$_{2}$O$_{7}$.
\end{abstract}


\maketitle 
%\tableofcontents

\section{Introduction} 
A variety of transition metal materials such as the cuprates \cite{Kivelson98Nature},
Ru-oxides \cite{Borzi07Science}, and Fe-pnictides \cite{Chu10Science} 
has been proposed to harbour an electronic nematic phase \cite{Fradkin10Review}.
Electronic nematic phases are broadly characterized by the presence of
spontaneously broken rotational symmetry and viewed as the quantum counterpart
of nematic classical liquid crystal phases.
The theoretical proposal of nematic quantum liquid crystals became more
concrete when experiments on ultra-pure bilayer ruthenate
(Sr$_{3}$Ru$_{2}$0$_{7}$) samples subjected to a magnetic field along the
c-axis revealed an unusual phase characterized by a pronounced residual
resistivity in place of a putative meta-magnetic 
quantum critical point (QCP) \cite{Grigera04Science}.
Interestingly, Sr$_{3}$Ru$_{2}$O$_{7}$ was initially
viewed as a prototype for the study of quantum phase
transitions, exhibiting a striking non-Fermi liquid resistivity
thought to originate from the putative magnetic field tuned QCP 
\cite{Grigera01Science}. 
The unusual phase found in ultra-pure samples is
delimited by two consecutive first order meta-magnetic
transitions at low temperature and, remarkably,
exhibits a significant in-plane magnetoresistive anisotropy
when the external field is slightly tilted towards one of the
in-plane crystal axes \cite{Borzi07Science}.
These observations strongly imply the formation of 
an anisotropic metallic, i.e. electronic nematic, phase in the
bilayer ruthenate compound.

Based at first on the two consecutive metamagnetic transitions, 
an electronic nematic phase was proposed
and generic features of nematic phase formation were 
theoretically explored early on \cite{Grigera04Science,Kee03PRB,Khavkine04PRB}. 
It was found that the transition between the isotropic and nematic
phase is generally first order, and that nematic order typically develops
near a van Hove singularity (vHS) to avoid a Lifshitz transition.
Varying the chemical potential, the nematic phase is
bounded at low and high values by two isotropic phases, while the 
concomitant first order transitions lead to jumps in the electron density.
When a magnetic field is applied (and the chemical potential is held
fixed at, say, some low value),
Zeeman coupling acts as a spin-dependent chemical potential term. 
Tuning the Zeeman field increases the volume of, e.g., the spin-up species 
which then enters the nematic phase near the vHS, resulting in a jump
in the spin-up density at the isotropic-nematic transition.
In contrast, the spin-down Fermi volume continuously decreases without
passing a van Hove point.
Such behavior gives rise to meta-magnetism, i.e. a sudden jump in the difference 
of up- to down-spin density at the isotropic-nematic transition \cite{Kee05PRB}.
Using more realistic band structures, microscopic models were
suggested and the above conclusion was found to be robust
\cite{Puetter07PRB,Puetter10PRB}.

Despite this mechanism of nematic phase formation, the origin of the critical
signatures remains mysterious
and recent thermodynamic 
data revealed additional complexity that requires new insight into 
the nematic theory \cite{Rost09Science}.
For over a decade in temperature $T$, the resistivity behaves nearly perfectly
linear \cite{Borzi07Science}, while the specific heat coefficient ($C/T$)
varies as $\text{log}(T)$ over the same range
\cite{Rost09Science,Rost11PNAS}. 
This behaviour resembles quantum criticality observed in a variety
of other correlated materials, suggesting the existence of an
unidentified QCP related to the nematic phase.
How can one reconcile the first order nature of the 
isotropic-nematic transition and the critical
behavior associated with a second order transition?

\begin{figure}[t!]
  \centering
  \includegraphics*[width=0.60\linewidth, clip]{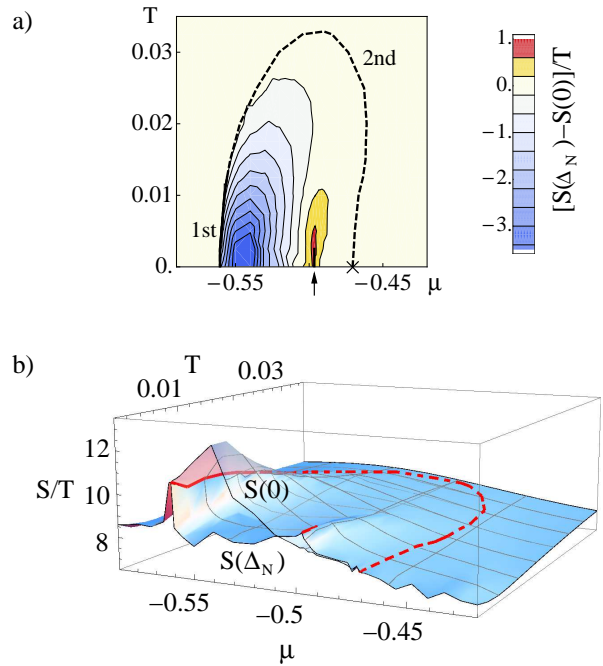}
  \caption{(a) Finite temperature phase diagram and contour plot of
    the entropy difference between the nematic and isotropic phase 
    $[S(\Delta_{\text{N}}) - S(0)]/T$. 
    The meta-nematic transition (marked by the arrow)
    and the QCP (marked by $\times$) are shown.
    (b) The corresponding absolute entropy landscapes 
    with ($S(\Delta_{\rm N})$) and without ($S(0)$) nematic ordering
    (for details see main text).
    \label{fig1}}
\end{figure}

In this paper, we present a  way to resolve this puzzle by
introducing two nematic order parameters relevant for multi-orbital
systems, and investigate as to how their interplay leads to interesting physics.
While the current work is motivated by the bilayer ruthenate, our
study on the isotropic-nematic transition is generally applicable 
to a t$_{\text{2g}}$-orbital system.
We show that in the presence of moderate spin-orbit (SO) coupling
the multi-orbital nature of the system is a key property for turning 
a nematic first order transition into a QCP.
We also show that the nematic QCP is accompanied by a nearby 
jump in nematicity, dubbed here meta-nematic transition, which 
obscures the QCP at low temperature.
This finding is in contrast to the current wisdom that
a QCP is hidden under the nematic dome.

Figure \ref{fig1} (a) shows the finite temperature phase diagram of such a 
multi-orbital system,
including the contour plot of the entropy difference between the nematic and 
the isotropic phase.
The nematic QCP marked by $\times$  
is found on the right low temperature phase boundary.
Note also that the entropy inside the nematic phase near the meta-nematic 
transition indicated by the arrow is higher than in the isotropic phase.
While the separation between the continuous and the meta-nematic 
transition depends on the underlying band structure and the SO coupling
strength, thermodynamics and transport at finite temperature 
will be governed by the QCP aside from singular density of states (DOS) effects, 
which may be relevant for the phenomena observed in Sr$_3$Ru$_2$O$_7$
\cite{Borzi07Science,Grigera04Science,Grigera01Science,Rost09Science,Perry01PRL,Kitagawa05PRL,Gegenwart06PRL}.
A more detailed discussion will be presented below.

\section{Nematic QCP}
\label{sec:II}
Theoretical progress has recently been made in developing microscopic 
models for nematic phase formation in Sr$_{3}$Ru$_{2}$O$_{7}$.
One approach suggested that the nematic phase is a spontaneous Fermi
surface (FS) distortion arising from a band with strong d$_{xy}$ character 
and a vHS near the Fermi level \cite{Puetter10PRB},
while another proposed that it originates from orbital ordering, i.e.,
the density difference between d$_{xz}$ and d$_{yz}$ orbitals, ignoring
the d$_{xy}$ orbital \cite{Raghu09PRB,Lee09PRB1}.
Both scenarios lead to broken x-y symmetry. 
To distinguish both types of ordering in the present paper, 
we call the former nematic and the latter orbital order.
Although the underlying microscopic driving mechanisms are distinct, 
both approaches have noted the importance of SO coupling, which is
in line with the
results of angle resolved photoemission spectroscopy (ARPES) 
\cite{Haverkort08PRL}.  
However, they failed to locate a QCP.
Thus, given that all t$_{\text{2g}}$ orbitals are coupled via SO interaction,
we explore here a broader notion of x-y symmetry breaking in a 
multi-orbital system.

To understand the origin of nematicity and its consequences, we start from
a tight-binding model, which reproduces the FS of
Sr$_{3}$Ru$_{2}$O$_{7}$ \cite{Tamai08PRL,Mercure09PRL,Mercure10PRB}
and was introduced in \cite{Puetter10PRB}.
The model is based on the Ru 4d t$_{2\text{g}}$ orbitals and includes
moderate SO coupling $H_{\text{SO}} = 2 \lambda \sum_{i}  {\bf
    L}_{i} \cdot {\bf S}_{i}$. 
We also incorporate a staggered lattice potential 
$H_{\text{st}} = \bar{g} \sum_{{\bf k}, \alpha, \sigma}
(c^{\alpha \dagger}_{{\bf k} \sigma} c^{\alpha}_{{\bf k}+{\bf Q}
  \sigma} + \text{h. c.})$ with ${\bf Q}=(\pi, \pi)$
to account for the effect of the staggered rotation of the RuO$_{6}$
octahedra, which doubles the real space unit cell 
\cite{Shaked00JSolidStateChem,Puetter10PRB,Fischer10PRB}.
In principle, the lattice potential can be momentum dependent
(e.g. $\bar{g}({\bf k}) \propto \text{cos}k_{x} + \text{cos} k_{y}$
as in \cite{Fischer10PRB})  but here we assume a constant 
value $\bar{g}$ as the results do not depend on its momentum 
structure.
The tight-binding band structure then has the form
\begin{eqnarray}
  H_{\text{0}} &=& \sum_{{\bf k}, \sigma=\uparrow, \downarrow} C^{\dagger}_{{\bf k} \sigma}
  \left(
    \begin{array}{ccc}
      A_{{\bf k} \sigma} & G \\
      G & A_{{\bf k}+{\bf Q} \sigma}
    \end{array}
  \right) 
  C_{{\bf k} \sigma}, \\
  A_{{\bf k} \sigma} &=&
  \left(
    \begin{array}{ccc}
      \varepsilon^{yz}_{{\bf k}} & \varepsilon^{1\text{D}}_{{\bf k}} +
      i \sigma \lambda & -\sigma \lambda \\
      \varepsilon^{1\text{D}}_{{\bf k}} - i \sigma \lambda & \varepsilon^{xz}_{{\bf
          k}} &  i \lambda \\
      -\sigma \lambda & -i \lambda &  \varepsilon^{xy}_{{\bf k}} 
    \end{array}
  \right), \; 
  G = \bar{g} {\bf 1}_{3\times3},
\end{eqnarray}
where $C^{\dagger}_{{\bf k} \sigma} = (c^{yz \dagger}_{{\bf k}
  \sigma}, 
c^{xz \dagger}_{{\bf k} \sigma}, c^{xy \dagger}_{{\bf k} \; -\sigma}, 
c^{yz \dagger}_{{\bf k}+{\bf Q} \sigma}, c^{xz \dagger}_{{\bf k}+{\bf
    Q} \sigma}, 
c^{xy \dagger}_{{\bf k}+{\bf Q} \; -\sigma})$
consists of electron operators creating an electron with spin $\sigma
= \uparrow, \downarrow$ in one of the t$_{\text{2g}}$-derived orbitals
$\alpha = yz, xz, xy$. 
The orbital dispersions are given by 
$\varepsilon^{yz/xz}_{{\bf k}} = -2 t_{1} \text{cos} k_{y/x} -2 t_{2}
\text{cos}k_{x/y} + \mu$, $\varepsilon^{xy}_{{\bf k}} = -2 t_{3}
(\text{cos}k_{x} + \text{cos}k_{y}) -4 t_{4} \text{cos}k_{x}
\text{cos}k_{y} -2 t_{5} (\text{cos}(2k_{x}) + \text{cos}(2 k_{y})) +
\mu$,
while the hybridization between the quasi-1D orbitals is 
$\varepsilon^{1\text{D}}_{{\bf k}} = -4 t_{6} \text{sin}k_{x} \text{sin} k_{y}$.
In the following the underlying band structure parameters are 
$t_{1}=0.5$, $t_{2}=0.05$, $t_{3}=0.5$, $t_{4}=0.1$, $t_{5}=-0.03$,
$t_{6}=0.05$, $\bar{g}=0.1$ and $\lambda=0.14$ (if not stated otherwise),
while the chemical potential $\mu$ serves as a tuning parameter.
All energies throughout the paper are expressed in units of 
the intra-orbital nearest neighbour hopping integral $2 t_{1}=1$. 

\begin{figure}[t!]
  \centering
  \includegraphics*[width=0.6\linewidth, clip]{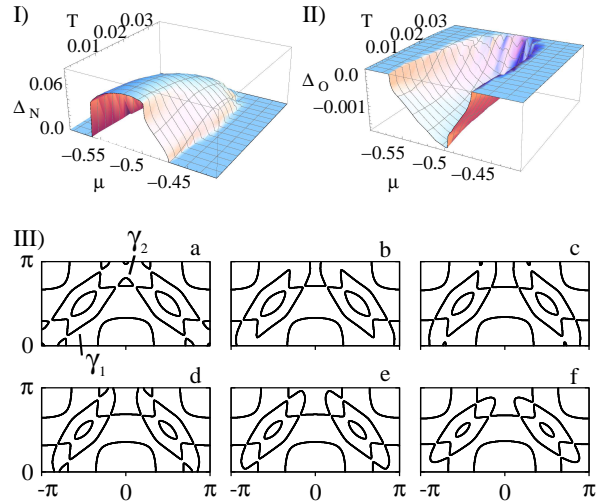}
  \caption{(I) \& (II) represent nematic 
    ($\Delta_{\text{N}}=\Delta^{\uparrow}_{\text N}=\Delta^{\downarrow}_{\text N}$) 
    and orbital ($\Delta_{\text{O}}$) order parameters over the $T$-$\mu$ plane, 
    respectively. Note the small magnitude of $\Delta_{\text{O}}$.
    (III) shows the qualitative changes of the FS in half of the original 
    Brillouin zone at selected points in the phase diagram for $T \approx 0$.
    In order of increasing $\mu$ these are taken 
    just before and after the nematic transition at
    $\mu_{\text{c1}}=-0.56$ (a \& b, respectively),
    at $\mu=-0.535$ (c), just before and after 
    the meta-nematic jump at $\mu_{\text{c2}}=-0.497$ (d \& e, respectively), 
    and at the continuous transition at $\mu_{\text{c3}}=-0.47$ (f).
    \label{fig2}}
\end{figure}
Since the nematic phase reported  in Sr$_3$Ru$_2$O$_7$ arises from a metallic 
state, it is reasonable to assume that long-range interactions 
are well screened, leaving moderately weak (compared with the bandwidth) 
on-site and nearest neighbour interactions.
The microscopic interaction Hamiltonian is then 
\begin{equation}
  \label{eq:Hint}
  H_{\text{int}} = U 
  \sum_{i, \alpha} n^{\alpha}_{i \uparrow} n^{\alpha}_{i \downarrow}
  + \frac{\tilde{U}}{2} \mathop{\sum_{i, \alpha \neq \beta}}_{\sigma, \sigma'} 
  n^{\alpha}_{i \sigma} n^{\beta}_{i \sigma'}
  + \mathop{\sum_{\langle i, j \rangle, \alpha}}_{\sigma, \sigma'} 
  V^{\alpha} n^{\alpha}_{i \sigma} n^{\alpha}_{j \sigma'},
\end{equation}
with $n^{\alpha}_{i \sigma} = c^{\alpha \dagger}_{i \sigma}
c^{\alpha}_{i \sigma}$ the density operator for electrons in 
orbital $\alpha$ at site $i$ and spin $\sigma$.
Here, $U$, $\tilde U$, and $V^{\alpha}$ represent repulsive intra-orbital
on-site, inter-orbital on-site, 
and intra-orbital nearest neighbour interactions, respectively.
We assume that $V^{\text {xz(yz)}}$ is finite along nearest neighbour x(y)-bonds, 
while $V^{\text {xy}}$ is finite along all four nearest neighbour bonds,
and that $V^{\alpha} \equiv V$. 

Different instabilities compete within $H_{0}+H_{\text{int}}$.
Most natural candidates are spin density wave and ferromagnetic instabilities.
However, it was shown in \cite{Raghu09PRB} that  
orbital ordering between d$_{yz}$ and d$_{xz}$ orbitals dominates  
when the inter-orbital interaction is significant
between the two quasi-1D orbitals in the presence of SO coupling. 
This important observation, however, did not take into account 
the 2D d$_{xy}$ orbital, which dominates the $\gamma$ 
FS sheets of Sr$_3$Ru$_2$O$_7$ as observed by ARPES \cite{Tamai08PRL}.  
In particular, the $\gamma_{2}$ sheet possesses a vHS
near the Fermi level implying that it is most 
susceptible to an instability.
Reference \cite{Puetter10PRB} indeed suggests that nematic ordering in
the xy dominated bands is the leading instability
when the nearest neighbour interaction is taken into account.
Since these theories imply that ferromagnetic and antiferromagnetic 
spin density wave instabilities are suppressed by SO coupling, we focus here
on the interplay of nematic and orbital ordering.

Introducing nematic and orbital order parameters,
$\Delta^{\sigma}_{\text{N}} = N^{-1} \sum_{{\bf k}} 
(\cos{k_x}-\cos{k_y}) \langle n^{xy}_{{\bf k} \sigma} \rangle $ 
and 
$ \Delta_{\text{O}} = N^{-1} \sum_{{\bf k}, \sigma} 
\langle n^{xz}_{{\bf k} \sigma} - n^{yz}_{{\bf k} \sigma} \rangle$,
respectively,
one arrives at the following Hamiltonian
\begin{eqnarray}
  \label{eq:MFH}
  H_{\text{MF}}  &=&   H_0
  - V_{\text{O}} \sum_{{\bf k}, \sigma} \big[
  \Delta_{\text{O}} (n^{xz}_{{\bf k} \sigma} - n^{yz}_{{\bf k} \sigma}) 
  + \frac{1}{4} \left( \Delta_{\text{O}} \right)^{2} \big]
  \nonumber
  \\
  && - V_{\text{N}} \sum_{{\bf k}, \sigma} \big[
  \Delta^{\sigma}_{\text{N}} (\cos{k_x}-\cos{k_y}) n^{xy}_{{\bf k} \sigma} 
  + \frac{1}{2}  \left(\Delta^{\sigma}_{\text{N}}\right)^{2} \big], 
\end{eqnarray}
where the effective interactions are given by 
$V_{\text{N}}=V$ and $V_{\text{O}} = \tilde{U}/2-U/4-V$. 
Note that the on-site intra-orbital interaction $U$ hinders orbital 
ordering (favouring magnetic ordering instead),
but does not interfere with nematic ordering. 
On the other hand, the inter-orbital interaction ${\tilde U}$ favours
orbital ordering, 
while the nearest neighbour intra-orbital interaction $V$ favours
nematic over orbital order.

\begin{figure}[t!]
  \centering
  \includegraphics*[width=0.6\linewidth, clip]{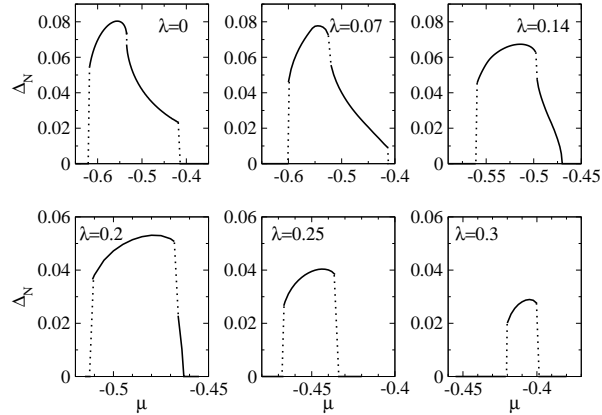}
  \caption{Each  panel represents the nematic order parameter 
    for various SO coupling strengths $\lambda$
    (dotted lines visualize discontinuities).
    Note that the first order transition  
    at the right phase boundary becomes continuous for
    $0.07 < \lambda < 0.2$.
    The results in figure \ref{fig1} and \ref{fig2}
    are based on $\lambda=0.14$.
    Very recent unpublished transport data  \cite{BruinUnpublished} in fact 
    bear a strong qualitative resemblance to this $\lambda=0.14$ plot.  
    The resistivity has a pronounced shoulder for fields higher than those of 
    the two first-order transitions, and within this shoulder becomes 
    anisotropic under the application of small in-plane magnetic fields.
    \label{DeltaN_SODependence}}
\end{figure}

The self-consistent solutions for 
nematic ($\Delta_{\text{N}} = \Delta^{\uparrow}_{\text{N}} = \Delta^{\downarrow}_{\text{N}}$) and orbital order over the $T-\mu$ plane are shown
in panels (I) and (II) of figure \ref{fig2}.
Panel (III) shows the FS at selected points for zero temperature.
As pointed out in figure \ref{fig1}, nematic and orbital order develop 
continuously from the isotropic phase at $\mu_{\text{c3}}=-0.47$,
in contrast to the generic first order transitions 
encountered when either nematic \cite{Kee03PRB,Khavkine04PRB,Puetter10PRB} 
or orbital \cite{Raghu09PRB} ordering is considered.

For the present discussion we have set $V_{\text{N}}=0.8$, while 
$V_{\text{O}} (\ll V_{\text{N}} )$ is chosen to be small or zero 
such that orbital order can only be induced by coupling to nematic 
order via SO interaction.
However, orbital ordering can also be induced for larger $V_{\text{O}}$ and 
smaller $V_{\text{N}}$ values than the above choice 
(for example for $V_{\text{O}}$ = 0.2 and $V_{\text{N}}$ =0.6) 
but does not alter the main conclusion of the existence of the 
QCP and the nearby meta-nematic jump. 
The reason for preferring the nematic to be the leading instability
($V_{\text{N}} > V_{\text{O}}$) is based on experimental constraints.
We found that an independent orbital instability occurs   
for $V_{\text{O}} \gtrsim 0.55$ (when $V_{\text{N}} =0$) but
that it is also accompanied by a drastic change in the overall FS topology,
exhibiting various strongly 1D-like open FS sheets
inconsistent with the de Haas van Alphen experimental
findings for Sr$_{3}$Ru$_{2}$O$_{7}$.
In contrast, the nematic instability gives rise to smooth changes of the
FS before and after the nematic window as shown in figure \ref{fig2} (III), 
in agreement with the de Haas van Alphen results 
\cite{Mercure09PRL,Mercure10PRB}.
Hence we consider the nematic instability as the relevant instability
in this study.
\footnote{While bilayer coupling may alter the role of 
  orbital and nematic orders, the existence of a nematic QCP
  is not sensitive to the choice of leading and induced orders.}

To understand the effect of SO coupling, we investigate how the 
phase transitions are modified by changes in the SO coupling strength
$\lambda$.
Figure \ref{DeltaN_SODependence} shows the nematic order parameter for 
different $\lambda$ 
(orbital order is not shown, but is small and disappears for $\lambda=0$).
The nematic order parameter exhibits typical first order 
transitions for small $\lambda$. 
For intermediate $\lambda$, the first order transition
at the higher critical chemical potential turns into a
continuous transition preceded by a sudden meta-nematic jump.
The change in nature from first order to second order is driven by the coupling 
between the two order parameters enabled by SO interaction
and can be captured qualitatively by the following Ginzburg-Landau (GL) free
energy analysis.

\section{GL free energy analysis} 
A generalized GL free energy takes the form 
\begin{eqnarray}
  {\cal F}_{\text{GL}} &=& \alpha_{\text{n}} \Delta_{\text{N}}^2 
  + \beta_\text{n} \Delta_{\text{N}}^4 + \gamma_{\text{n}} \Delta_{\text{N}}^6  
  + \alpha_{\text{o}} \Delta_{\text{O}}^2 + \beta_{\text{o}}
  \Delta_{\text{O}}^4 + \gamma_{\text{o}} \Delta_{\text{O}}^6 
  \nonumber \\
  &&+ \delta \Delta_{\text{N}} \Delta_{\text{O}} [1 + \eta
  (\Delta_{\text{N}}^2 + \Delta_{\text{O}}^2)] + \kappa 
  \Delta_{\text{N}}^2 \Delta_{\text{O}}^2 + \dots,
\end{eqnarray}
where we assume that the higher order terms denoted by the 
ellipsis are negligible. 
The GL free energy analysis is a phenomenological approach, and it is important to understand how each term is 
allowed by symmetry, in particular the $\delta$ term.
The nematic order parameter breaks only the x-y symmetry of the underlying lattice (or the rotational symmetry in the case of the Jellium model).
Thus the GL free energy should be invariant under a sign change of the
order parameter (e.g. $\Delta_{\text{N}} \rightarrow -\Delta_{\text{N}}$, 
corresponding to a $\pi/2$ rotation), which restricts the GL expansion to only even powers in the nematic order parameter. 
Using the Jellium model of the electron gas, the GL free energy with 
nematic order was analyzed in \cite{DellAnna07PRL}, 
where the isotropic-nematic transition is assumed to be
continuous with $\alpha_{\text{n}}$ changing sign at the transition, 
while $\beta_{\text{n}}$ stays positive.
However, it was first pointed out in \cite{Khavkine04PRB} using a
single nematic order parameter that
the free energy has a log-singularity due to a vHS (Lifshitz
transition) leading to a first order isotropic-nematic transition
in a simple square lattice model. 
An order parameter expansion shows that the $\beta_{\text{n}}$ coefficient
becomes negative while $\alpha_{\text{n}}$ and $\gamma_{\text{n}}$ remain
positive at the transition point \cite{Khavkine04PRB}.
Since then, turning the first order transition into second order has
become an interesting problem.
In the present study of multi-orbital systems such as 
the t$_{\text{2g}}$-orbital Sr$_{3}$Ru$_{2}$O$_{7}$ compound,
there are two x-y symmetry breaking order parameters as introduced in the previous section.
Both orbital ordering (i.e. the density imbalance between d$_{xz}$ and d$_{yz}$ orbitals) 
and nematic ordering (defined on the d$_{xy}$-orbital manifold) break
the same x-y symmetry. 
In principle, the GL free energy contains a finite coupling term 
such as $\sim \Delta_{\text{N}} \cdot \Delta_{\text{O}}$, as long as
the Hamiltonian contains
mixing between d$_{xy}$ and quasi-1D (d$_{xz}$ and d$_{yz}$) orbitals. 
However, due to the nature of the orbitals and the inversion symmetry of the  square lattice about the x-y plane, 
finite hopping integrals between d$_{xy}$ and the quasi-1D orbitals are absent. 
Therefore, the dominant mixing between d$_{xy}$ and quasi-1D orbitals is SO coupling.

Under the symmetry considerations discussed above,
the $\delta$-term is allowed when 
SO interaction is present, i.e.  $\delta \propto \lambda + O(\lambda^{3})$.
Starting with positive coefficients only, no ordering is initially present.
However, as was shown previously for a single band, the
nematic instability is typically
of first order due to $\beta_{\text{n}}<0$ and $\alpha_{\text{n}},
\gamma_{\text{n}}>0$ \cite{Khavkine04PRB}.
Therefore, without SO coupling ($\delta=0$) and 
$\alpha_{\text{o}}, \beta_{\text{o}}>0$ finite nematic order
develops when $\beta_{\text{n}}^2 > 4 \alpha_{\text{n}} \gamma_{\text{n}} $, 
while orbital ordering is absent ($\Delta_{\text{O}}=0$).
On the other hand, when $\delta \neq 0$ nematic order starts to develop
continuously for $\delta^2  > 4 \alpha_{\text{n}} \alpha_{\text{o}}$, inducing
also orbital ordering (with $\Delta_{\text{O}} = - \frac{\delta}{2
  \alpha_{\text{o}}} \Delta_{\text{N}}$ up to second order GL terms in
$\Delta_{\text{O}}$).
A jump in the order parameters furthermore occurs when, e.g., the
quartic order $\eta$-term becomes sufficiently negative.

\section{DOS, entropy and finite temperature phase diagram}
The finite temperature phase diagram is displayed in figure \ref{fig1} (a).
The phase boundary between the isotropic ground state and 
the nematic phase is continuous except near $\mu_{\text{c}1} = -0.56$,
where a first order transition appears,
extending up to $T \approx 0.015$ in temperature.
Furthermore, the meta-nematic transition appearing at low temperature near
$\mu_{\text{c}2} = -0.497$ disappears at a slightly higher temperature.
Remarkably, the slope of the phase boundary near the QCP at 
$\mu_{\text{c}3} = -0.47$ is positive, indicating that the nematic phase 
has a higher entropy than the isotropic phase.

\begin{figure}[t!]
  \centering
  \includegraphics*[width=0.80\linewidth, clip]{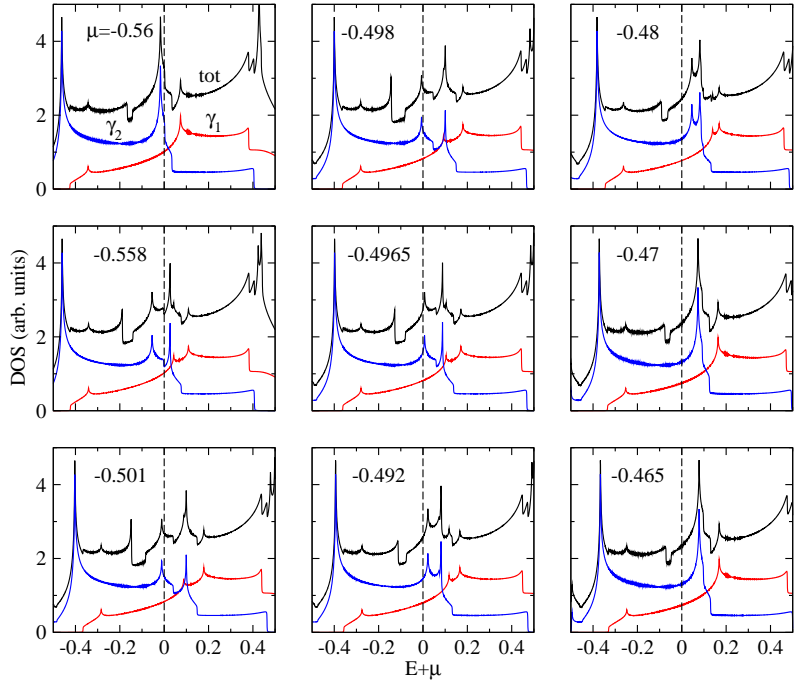}
  \caption{Evolution of the DOS of all bands combined (black
      curves) and of the $\gamma_{1}$ (red) and the $\gamma_{2}$
      (blue) band (contributing most to the DOS near the Fermi level
      which is indicated by the vertical dashed line) with the 
      chemical potential $\mu$ at $T=0$ (see also figures 
      \ref{fig1} and \ref{fig2}). Note the splitting of the $\gamma$
      band van Hove peaks near the Fermi level inside the nematic phase.
    \label{fig4}}
\end{figure}

To check this, we calculated the entropy
\begin{equation}
  S = -\sum_{n} \sum_{\bf k} \left[ f(E^n_{\bf k}) \;
    \text{ln}(f(E^n_{\bf k})) + (1- f(E^n_{\bf k})) \;  
    \text{ln}(1-f(E^n_{\bf k})) \right],
\end{equation}
where $f(E)$ is the Fermi function, which is also shown in figure \ref{fig1}.
Over most of the region the entropy associated with the bare band structure
is larger than the one for the nematic phase
and peaks (or, rather, logarithmically diverges) at low temperature 
near $\mu=-0.54$ due to the vHS in the DOS
associated with the saddle points in the $\gamma_{2}$ band as shown in 
figure \ref{fig1} (b).
This is expected since previous work \cite{Khavkine04PRB,Puetter07PRB}, 
as discussed in the introduction, showed
that the formation of the nematic phase helps to avoid 
a large DOS near the Fermi level, and hence a large entropy,
by splitting a logarithmic DOS peak into two peaks 
located further away from the Fermi energy.
One of the peaks is indeed shifted to lower $\mu$ values in the
nematic phase, which is illustrated in figure \ref{fig4}
(see e.g. the $\gamma_{2}$ peak(s) in the panels 
for $\mu=-0.56$ and $-0.558$ near the first order
transition at $\mu_{\text{c1}}$).
Note also that the DOS peak deriving from $\gamma_{1}$ band near the
Fermi level clearly reacts to the formation of nematic order by splitting as well.

There is another peak in the entropy landscape
at the meta-nematic transition $\mu_{\text{c2}}$ inside the nematic phase 
where the nematic entropy exceeds the entropy of the isotropic phase.
This singular behaviour is unexpected and occurs since finite nematicity persists
beyond $\mu_{\text{c2}}$ due to the coupling of nematic and 
orbital order as discussed in GL free energy analysis.
This unusual feature is also responsible for the concave
curvature of the phase boundary in the finite temperature phase diagram 
in figure \ref{fig1} (a).
In the DOS the meta-nematic transition corresponds to a sudden
changeover from one side of the Fermi level to the other of the lower of the two split
$\gamma_{2}$ van Hove peaks (cf. the sequence from $\mu=-0.498$ to $-0.492$ 
in figure \ref{fig4}), while only at the QCP ($\mu_{c3}=-0.47$) do 
the $\gamma$ vHS merge into one (respectively for $\gamma_{1}$ 
and $\gamma_{2}$).

\section{Discussion and summary}
Quantitatively, the present model is based off the bilayer 
ruthenate compound, where 
the formation of a nematic phase is driven by a moderate magnetic field 
applied along the c-axis \cite{Borzi07Science}.
Although one of the tuning parameters of the present phase diagram is 
the chemical potential, let us attempt to relate our results to 
the experimental, magnetic field tuned phase diagram \cite{Rost09Science}.
Within the weak coupling approach, the leading effect of a magnetic
field along c-axis is to shift the location (in momentum and energy)  
of the relevant van Hove points in the $\gamma$ bands
due to the Zeeman term \cite{Puetter10PRB}.
In the absence of SO coupling, the Zeeman term largely acts as a
uniform (in momentum space),
spin-dependent chemical potential as discussed in section \ref{sec:II}.
In the presence of SO coupling, which mixes up- and down-spins, 
the effective chemical potential shift becomes momentum dependent.
The shift is not uniform but determined by the spin composition of each band at 
a given momentum ${\bf k}$. 
As discussed, the relevant $\gamma_{2}$ and $\gamma_{1}$ vHS arise from
a small region near $(\pm \pi, 0)$ and $(0, \pm \pi)$ 
in momentum space with a large d$_{xy}$ admixture.
The effective shift of the $\gamma_{i s}$ band (where $s=\pm$ 
denotes the pseudospin and $i=1,2$) 
therefore is determined by the spin nature of
the contributing d$_{xy}$ orbital (i.e. either c$^{xy}_{{\bf k} \uparrow}$ or
c$^{xy}_{{\bf k} \downarrow}$). 
Hence we do not expect qualitative changes to our main conclusion, 
i.e. the existence of a QCP and a meta-nematic transition, when the
tuning parameter is an external magnetic field oriented along the c-axis.

However, one may expect that for a tilted field the interplay of
SO interaction, Zeeman term, and additional orbital couplings may 
affect the underlying band structure in a complicated manner.
The effect of an arbitrary magnetic field on nematic, orbital and
magnetic instabilities therefore is complex and requires a
self-consistent theory with a number of different order parameters including 
independent spin-up and -down nematic and orbital orderings and 
orbital dependent magnetization channels, so that the present 
result has limited applicability.

Moreover, the phase boundary at low $\mu$ and low $T$ bends towards the nematic
phase, opposite to the experimental observations.
However, additional bilayer effects possibly 
change the underlying band structure such that the curvature becomes
convex due to DOS effects and may even give rise to an additional nematic QCP.
Further studies of the interplay of nematicity, orbital ordering, 
bilayer coupling, magnetization, and SO coupling in the presence of a 
magnetic field (accounting for both orbital and spin effects) are
highly desirable and will be presented elsewhere \cite{PuetterFuture}.

Naturally, critical fluctuations accompanying 
the nematic QCP give rise to a quantum critical fan at finite temperature. 
Considering DOS and band structure effects only, one can generally expect 
that thermodynamic quantities diverge at most logarithmically.
In contrast, the experimental specific heat and entropy curves 
for Sr$_{3}$Ru$_{2}$O$_{7}$ follow a power law $\propto H_{c}/(H-H_{\text{c}})$ 
as a function of magnetic field $H$ when approaching the nematic phase,
indicating that critical fluctuations are important
\cite{Rost09Science,Rost10PSSb} and substantiating our finding.
In addition to entropy, 
critical non-Fermi liquid response was
observed early on in resistivity measurements when 
approaching the nematic regime from finite temperature
\cite{Grigera01Science},
exhibiting a $\sim T^{\alpha}$ dependence with $\alpha \approx 1.0-1.5$.
Theoretically, nematic quantum critical fluctuations in continuum
and lattice systems have been investigated as well
and shown to lead to non-Fermi liquid behaviour
\cite{Oganesyan01PRB,Metzner03PRL,Kao07PRB,DellAnna07PRL}.
However, a critical fluctuation theory in the presence of a singular
DOS and in particular the magneto-thermal critical behaviour 
in close proximity to a meta-nematic transition at low temperature is still missing. 
It is also worthwhile to note that  magnetic susceptibility measurements will be most 
sensitive to the divergences at the first order transition and the meta-nematic 
transition inside the nematic phase, which obscures the QCP.
Such a study lies beyond the scope of the present paper and will be 
addressed in the future.
While the details of the current results are based on the band
dispersion of the bilayer ruthenate, our GL analysis with multiple
order parameters, describing both continuous and disguised first order 
(jump in the order parameters) transitions, is in principle 
applicable to a range of quantum phase transitions in itinerant 
multi-orbital materials.

In summary,  we have shown that SO coupling in a multi-orbital 
t$_{2\text{g}}$ system can have a considerable impact on the formation
of a nematic phase. 
Although the nematic instability arises from the d$_{xy}$ band, it 
also induces a density imbalance between d$_{yz}$ and d$_{xz}$ orbitals in the 
presence of SO interaction, which in turn 
changes the nematic phase transition from first to second order,
such that the FS topology remains qualitatively 
unchanged across the transition.
Under what circumstances 
a continuous or a first order phase boundary 
is more favourable in the presence of SO coupling
likely depends on the detailed band structure and 
energetics as discussed in our analysis of the GL energy.
Furthermore, since the continuous transition
does not preempt a vHS, which in the present case happens 
via a nearby meta-nematic transition, the entropy diverges inside the
nematic phase, causing a concave curvature of the phase boundary.
The meta-nematic transition also hinders the observation of 
the QCP at low temperature, making further experimental analysis of 
the thermodynamics near the nematic phase boundaries 
in Sr$_{3}$Ru$_2$O$_7$  necessary to confirm the present 
theoretical proposal.

\ack
\addcontentsline{toc}{section}{Acknowledgements}
We thank A.~P. Mackenzie for collaborative discussions via 
the Canadian Institute for Advanced Research.
This work was supported by NSERC of Canada and Canada Research Chair.

\end{document}